\documentclass[preprint2]{aastex}
\input psfig.sty

\newcommand{\sax}{{\it BeppoSAX}}
\newcommand{\xte}{{\it RXTE}}

\newcommand{\asca}{{\it ASCA}}
\newcommand{\ginga}{{\it Ginga}}
\newcommand{\gro}{{\it CGRO}}

\newcommand{\ergcms}{\mbox{erg cm$^{-2}$ s$^{-1}$}}

\begin{document}
       
\title{Broad band spectrum of Cygnus X-1 in two spectral states
with BeppoSAX}

\author{F. Frontera\altaffilmark{1,2},
E. Palazzi\altaffilmark{2},
A. A. Zdziarski\altaffilmark{3},
F. Haardt \altaffilmark{4},
G. C.~Perola\altaffilmark{5},
L.~Chiappetti\altaffilmark{6},
G.~Cusumano\altaffilmark{7},
D.~Dal Fiume\altaffilmark{2},
S.~Del Sordo\altaffilmark{7},
M.~Orlandini\altaffilmark{2},
A. N.~Parmar\altaffilmark{8},
L.~Piro\altaffilmark{9},
A.~Santangelo\altaffilmark{7},
A.~Segreto\altaffilmark{7},
A.~Treves\altaffilmark{4}, and
M.~Trifoglio\altaffilmark{2}
}

\altaffiltext{1}{Dipartimento di Fisica, Universit\`a degli Studi di Ferrara,
Via Paradiso 12, I-44100 Ferrara, Italy; frontera@fe.infn.it}

\altaffiltext{2}{Istituto Tecnologie e Studio Radiazioni Extraterrestri, CNR,
Via Gobetti 101, 40129 Bologna, Italy}

\altaffiltext{3}{N. Copernicus Astronomical Center, Bartycka 18,
00-716  Warsaw, Poland; aaz@camk.edu.pl}

\altaffiltext{4}{Universit\`a dell'Insubria, Via Lucini 3, I-22100
Como, Italy}

\altaffiltext{5}{Dipartimento di Fisica ``E. Amaldi'', Universit\`a degli
Studi ``Roma Tre'',
Via della Vasca Navale 84, I-00146 Roma, Italy}

\altaffiltext{6}{Istituto di Fisica Cosmica ``G. Occhialini'', CNR,
 Via Bassini 15, I-20133 Milano, Italy}

\altaffiltext{7}{Istituto di Fisica Cosmica ed Applicazioni dell'Informatica,
CNR,
Via U. La Malfa 153, I-90146 Palermo, Italy}

\altaffiltext{8}{Astrophysics Division, Space Science Department of ESA,
2200 AG Noordwijk, The Netherlands}


\altaffiltext{9}{Istituto di Astrofisica Spaziale, CNR,
Via Fosso del Cavaliere, I-00133 Roma, Italy}

\begin{abstract}
We report on the 0.5--200 keV spectral properties of Cyg X-1 observed at
different epochs with the Narrow Field Instruments of the \sax\/ satellite. The
source was in its soft state during the first observation of 1996 June. In the
second observation of 1996 September, the source had parameters characteristic
to its hard state. A soft X-ray excess, a broad Fe K$\alpha$ line and Compton
reflection are clearly detected in both states. The soft-state broad-band
continuum is well modeled by a disk blackbody (accounting for the soft excess)
and Compton upscattering of the disk photons by a hybrid, thermal/non-thermal,
plasma, probably forming a corona above the disk (also giving rise to the
Compton-reflection component). In the hard state, the primary hard X-ray
spectrum can be well modeled by Compton upscattering of a weak blackbody
emission by a thermal plasma at a temperature of $\sim 60$ keV. The soft excess
is then explained by thermal Comptonization of the same blackbody emission by
another hot plasma cloud characterized by a low value of its Compton parameter.
Finally, we find the characteristic ratio of the bolometric flux in the soft
state to that in the hard state to be about 3. This value is much more
compatible with theories of state transitions than the previously reported 
(and likely underestimated) value of 1.5.
\end{abstract}


\section {Introduction}
\label{intro}

Cyg X-1 is one of the brightest and most extensively studied X-ray sources
in the sky. Its optical companion is the O9.7 Iab supergiant HDE 226868.
Estimates of the mass, $M$, of the X-ray star, $5\la M/{\rm M}_\odot\la 15$
(e.g., Herrero et al.\ 1995), strongly suggest the presence of a black hole.

Its distance has been claimed to be $d\ga 2.5$ kpc based on comparison of the
extinction of Cyg X-1, $E(B-V)\sim 1.0$--1.1, with that of field stars (Bregman
et al.\ 1973; Margon, Bowyer \& Stone 1973). On the other hand, HDE 226868
belongs, most likely, to the NGC 6871/OB3 association (Massey, Johnson \&
DeGioia-Eastwood 1995), in which case a large fraction of the extinction is
local to the association. The distance to the association was measured as
$d\approx 1.8\pm 0.5$ kpc (Janes \& Adler 1982), $2.1\pm 0.1$ kpc (Massey,
Johnson \& DeGioia-Eastwood 1995), and  1.8 kpc (Malysheva 1997). Here, we
adopt $d=2$ kpc, which is also the value assumed by Gierli\'nski et al.\ (1999,
hereafter G99).

On time scales from weeks to years, the 2--10 keV flux typically shows two
intensity states: a low state, in which the source spends $\sim 90 \%$ of the
time, characterized by a luminosity of $L(2$--10 keV) $\sim 3 \times
10^{36}$ erg s$^{-1}$ and by a hard power-law spectrum ($\propto E^{-\Gamma}$)
with a photon spectral index of $\Gamma \sim 1.7$ (e.g., Gierli\'nski et al.
1997, hereafter G97), and  a high state corresponding to  a 2--10~keV
luminosity an order of magnitude higher, with a strong blackbody component with
$kT\sim 0.5$ keV and soft power law tail with $\Gamma\sim 2$--3 (Dolan et al.\
1977; Ogawara et al.\ 1982). Based on these spectral properties, we will
hereafter call the states hard (HS) and soft (SS). The low-energy ($\la 10$
keV) and the high-energy ($\ga 20$ keV) fluxes are anticorrelated in the two
states (see review by Liang \& Nolan 1984), while the bolometric luminosity
changes weakly (Nowak 1995; Zhang et al.\ 1997a).

Based on the shape of the high-energy cutoff to the power-law spectrum in the
HS, the dominant radiative process appears to be thermal Comptonization
(Sunyaev \& Tr\"umper 1979; Nolan et al.\ 1981; Phlips et al.\ 1996; G97). The
seed photons for this process are, most likely, provided by blackbody emission
of some cold medium, a component of which is also seen in the HS at low
energies (e.g., Ebisawa et al.\ 1996, hereafter E96). The cold medium probably
forms an accretion disk, as indicated by the presence of Compton reflection
(Done et al.\ 1992; G97) and a fluorescent Fe K$\alpha$ emission (Done et al.\
1992; E96; G97).

\begin{figure}
\epsscale{1.0}
\plotone{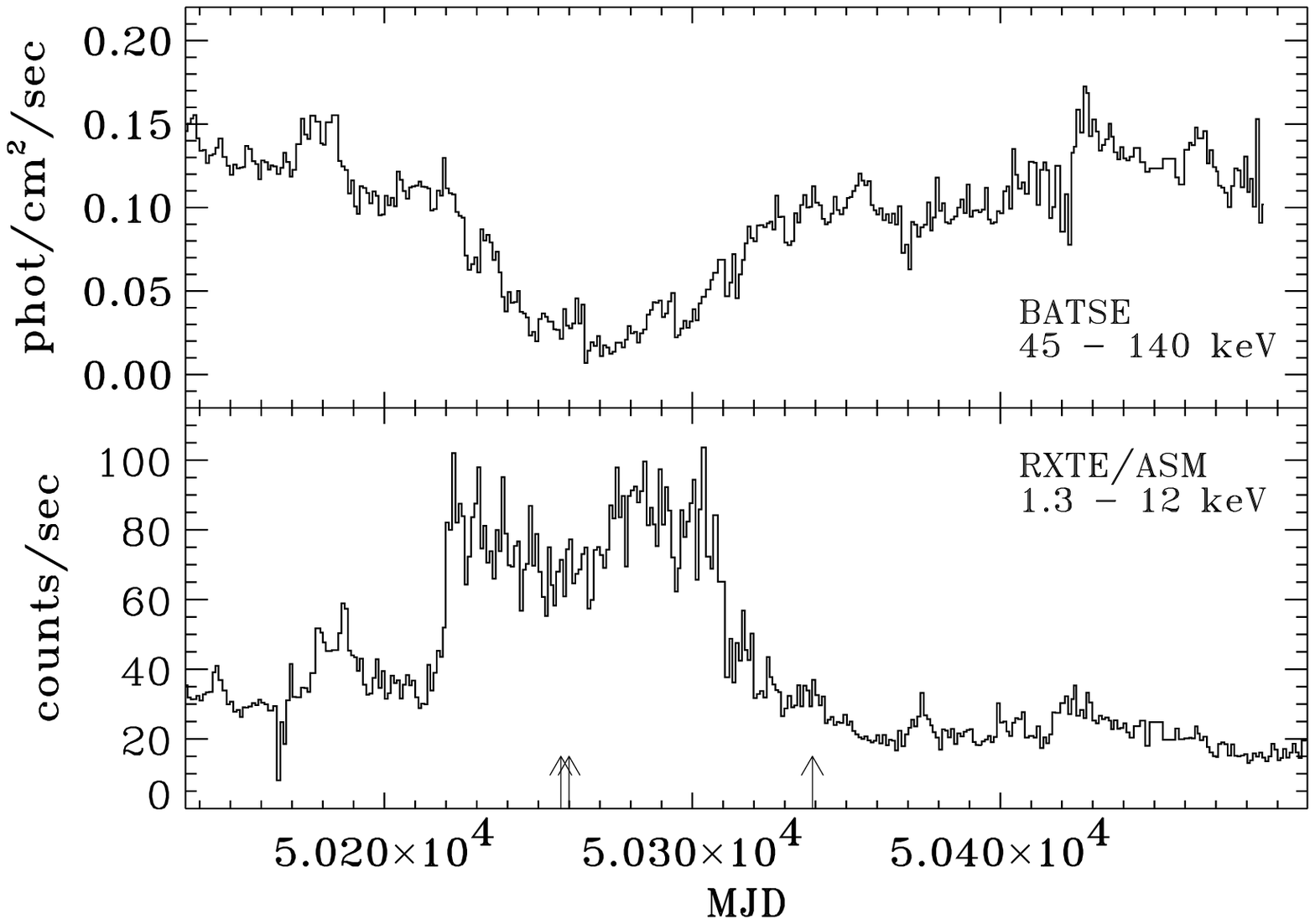}
\caption{One-day-average light curve of Cyg X-1 detected with the {\it
CGRO}/BATSE and \xte/ASM during the 1996 state transitions. (From the public
archives at cossc.gsfc.nasa.gov/batse/hilev/occ.html and
xte.mit.edu/XTE/asmlc/ASM.html, respectively.) The \sax\/ observations are
indicated with arrows.
}
\label{f:transition}
\end{figure}

In 1996 May, Cyg X-1  underwent a transition (see Fig.\ 1) from the HS to the
SS, with an increase of the 2--10 keV flux \cite{Cui96} and a decrease of the
20--200~keV flux \cite{Zhang96}. The SS lasted until 1996 September,
and, during that period, several X-ray observations were performed. Belloni et
al.\ (1996) report on observations with \xte/PCA on May 22, 23 and 30, and one
in 1996 February, when the source was in its normal HS. The May spectra were
fit with a `multicolor' blackbody model \cite{Mitsuda84} with  $kT_{\rm in} =
0.36 \pm 0.01$ keV at the inner radius of the disk and a power law with
$\Gamma=2.15\pm 0.02$. The February spectrum was fit with a power-law with
$\Gamma=1.60\pm 0.02$, a Fe K line and an edge. Dotani et al.\ (1997) reported
results of an \asca\/ observation (0.5--10 keV) performed on 1996 May 30--31.
The lower energy part of the spectrum (0.5--4.5 keV) was fit by either the
`multicolor' disk model with inner temperature of $kT_{\rm in} = 0.43$ keV or
by a general-relativistic blackbody disk model \cite{Hanawa89}. The higher
energy spectrum (4.5--10 keV) was accounted for a power law ($\Gamma= 2.3\pm
0.1$) and a `smeared edge' model. Combining the \asca\/ observation with the
simultaneous \xte\/ observation, Cui et al.\ (1998) fit the 0.7--50 keV
spectrum with a model consisting of `multicolor' blackbody, thermal
Comptonization with electrons at $kT\simeq 40$ keV, Compton reflection, and a
broad Fe K$\alpha$ line feature (with a Gaussian width of $\sim 0.35$ keV and
an equivalent width  of $\sim 130$ eV). However, physical interpretation of the
results is rather unclear because the temperature of the seed photons for
Comptonization was more than twice the maximum disk blackbody temperature.

The simultaneous \asca/\xte\/ data were also considered by G99. They found
that thermal Comptonization at 40 keV (as obtained by Cui et al.\ 1998) was
unable to reproduce the \gro/OSSE data (of 1996 June 14--25), with a power law
with $\Gamma\sim 2.5$ extending to $\ga 600$ keV without a cutoff. An increase
of the plasma temperature could improve the fit at high energies but would
strongly worsen the disagreement with the data below $\sim 10$ keV (see also
Coppi 1999). Furthermore, thermal-Compton models in which the source of seed
photons for upscattering was the disk emission were completely unacceptable
even for the \xte/\asca\/ data alone. Thus, G99 concluded that the
thermal-Compton model for the SS can be rejected.

An alternative model found by G99 to fit well the \asca/\xte/OSSE data was that
of a non-thermal plasma (with electron acceleration) located in the vicinity of
the accretion disk, possibly forming a corona. In steady state, the electron
distribution is hybrid, containing both thermal and non-thermal components, and
both components Compton upscatter the disk emission. The specific model is
developed from a code by Coppi (1992).

In this work, we study two sets of observations of Cyg X-1 performed by \sax\/
in 1996. During the first set, the source was in the SS. During the second one,
it was already back in its HS, based on its spectral properties as determined
by the {\it RXTE}/ASM (Wen, Cui \& Bradt 2000). The X-ray spectrum during that
observation is softer than that measured by G97, but similarly soft spectra are
commonly seen by the ASM during the HS. In fact, our second observation
corresponds to about the middle of the region covered by the ASM data on a
color/intensity diagram during the hard state (Zdziarski, Wen \& Paciesas, in
preparation). Furthermore, it appears to fall after the periods of enhanced
time lags characteristic to state transitions in Cyg X-1 (Pottschmidt et al.\
2000).

Unlike other missions, \sax\/ offers the possibility to cover a wide energy
band (0.1--300 keV), crucial to constrain models. Moreover, we can compare the
spectra obtained with the same instruments in different states.  Preliminary
results of spectral analysis of the observations were reported by Palazzi et
al.\ (1999). In \S \ref{obs} below, we describe the observations and discuss
time variability of our data. In \S \ref{analysis}, we describe our method of
spectral analysis, and its results are given in \S \ref{results}. Finally, in
\S \ref{discussion}, astrophysical consequences of our results are discussed.

\section{Observations, Data Sets and Timing}
\label{obs}

The \sax\/ payload includes 4 Narrow Field Instruments (NFIs, Boella et al.\
1997a), which embody a Low Energy Concentrator Spectrometer (LECS, 0.1--10 keV,
Parmar et al.\ 1997), 3 Medium Energy Concentrators Spectrometers (MECS,
1.3--10 keV, Boella et al.\ 1997b), a High Pressure Gas Scintillator
Proportional Counter (HPGSPC, 3--100 keV, Manzo et al.\ 1997), and a Phoswich
Detection System (PDS, 15--300 keV, Frontera et al.\ 1997). Both LECS and MECS
have imaging capabilities, while the HPGSPC and PDS are direct-viewing
detectors with a Field of View (FOV) of $1\degr$ and $1.3\degr$, respectively.
The latter instruments use rocking collimators for background monitoring.

\begin{deluxetable}{l c c c c c c}
\tabletypesize{\small}
\tablewidth{0pt}
\tablecaption{The observation log}
\tablehead{
\colhead{obs\#}    & \colhead{Date}    & \colhead{Start time} & \colhead{End
time}  &\colhead{Orbital phase}  & \colhead{Instrument} & \colhead{Exposure
[s]} }
\startdata
1      & 1996 June 22 & 18:33:21   & 20:05:18    &  0.93--0.94 & LECS  & 844 \\
       &              &            &             &             &
MECS\tablenotemark{a}  & 2328   \\
       &              &            &             &        & HPGSPC   & 750   \\
&              &            &             &             & PDS      &  1480  \\
2      & 1996 June 25 & 09:03:15   & 13:40:05    & 0.40--0.43  & MECS     &
7461  \\
3A     & 1996 Sept 12 & 00:52:00   & 11:09:50    & 0.51--0.52  &  LECS    &
4641  \\
&              &            &             &             & MECS     &   19653 \\
&              &            &             &             &
HPGSPC\tablenotemark{b}   &   8845  \\
       &              &            &             &             &
PDS\tablenotemark{b}      &   10560 \\
3B     & 1996 Sept 12 & 11:10:11   & 15:44:50    & 0.52--0.55  & LECS     &
150  \\
       &              &            &             &             & MECS     &
8915 \\
       &              &            &             &             & HPGSPC   &
7863 \\
       &              &            &             &             & PDS      &
2844 \\
\enddata
\tablenotetext{a}{Unusable due to the offset of $14'$ of the source
from the instrument axis.}
\tablenotetext{b}{Unusable due to no available background.}

\end{deluxetable}

The first Cyg X-1 observation ({\it obs\#1}) was performed on 1996 June 22
(first light of \sax). The source was offset by about $14'$ with respect to the
instrument axes, as established from the MECS image. The source was observed
again on-axis ({\it obs\#2}) on 1996 June 25. Unfortunately, in this
observation, only the MECS were  switched on. Other observations were performed
in 1996 September 12--13 with all the NFIs operative. These observations were
devoted to a diagonal scan of the NFI field of view to assess their off-axis
response. Two of them  ({\it obs\#3A, B}) were almost on axis (offset $<3'$).
The log of the observations is given in Table 1, together with the phase of the
5.6~d binary orbit with respect to the superior conjunction of the X-ray source
(using the ephemeris of Brocksopp et al.\ 1999).

Useful data from {\it obs\#1, obs\#2\/} and {\it obs\#3} were selected from
time intervals that met the following criteria: satellite outside the South
Atlantic Geomagnetic Anomaly, the elevation angle above the Earth limb
of $\geq 5\degr$, dark Earth (for LECS), stabilized high voltage supplies. For
{\it obs\#1}, the MECS data were not usable for spectral analysis because the
centroid of the image coincided with the strong-back of the detector entrance
window, which introduced a complex spatial modulation of the window
transparency.

The LECS and MECS source spectra were extracted from regions with radii of $8'$
and $7'$, respectively, around the centroid of the source image. We used as
background spectrum that obtained from the observation of blank fields. The
spectra from the 3 MECS units were equalized and co-added. For {\it obs\#2}, we
analyzed only data from MECS units 2 and 3 since the data files of the unit 1
at the epoch of this observation were corrupted. The  background levels of
HPGSPC and PDS were estimated by swapping their collimators every 96 s. During
{\it obs\#3A}, the detectors were continuously pointed at the source, so the
background level was not available for this observation. Therefore, for the
broad-band spectral analysis we used {\it obs\#3B}, during which the background
was continuously monitored.

Although our paper deals primarily with spectral analysis, we have also
performed some temporal analysis of our data. Specifically, we have obtained
the cross-correlation function between the lightcurves in two bands of the MECS
detector, soft, 1.3--4 keV, and hard, 4--10 keV, in the 2 spectral states. The
cross-correlation was computed using an FFT algorithm (under {\sc xronos}) for
the time series binned at the time resolution of 100, 10 and 1 ms. We have
found that the cross-correlation function does not depend on the binning, and
we have detected no measurable time lags, in agreement with the results of an
analysis of the \xte/PCA data by Maccarone, Coppi \& Poutanen (2000). Note that
the \sax\/ detectors are much less sensitive than those of the \xte/PCA, which
remains the primary tool for this kind of analysis.

Then, we have considered the possible spectral variability of the source. We
have found that the light curves integrated over 200 s show a source time
variability up to $\sim 30\%$ on time scale of several hundred seconds. We have
tested for {\it obs\#3A\/} the spectral variability  by subdividing that
observation into 3 segments of $\sim 10^4$ s each. No statistically significant
variation of the best-fit parameters from one segment to another was found.
Thus, below we consider the spectra averaged over each observation.

\section {Spectral Analysis}
\label{analysis}

The count rate spectra  were  analyzed using the {\sc xspec} v.\ 10 software
package (Arnaud 1996).  The adopted response matrices took into account the
offset, if any, of the observed source centroid. Intercalibration analysis of
the \sax\/ NFIs using Crab Nebula (Fiore, Guainazzi, \& Grandi 1999) has shown
that systematic uncertainties in the spectrum determination  are $\sim 1.5\%$.
In order to limit these systematic errors in the deconvolved spectra, the
energy band of each instrument was limited to that where the response function
was best known: LECS, 0.5--4~keV; MECS, 3--10~keV; HPGSPC, 5--40~keV; PDS,
15--200~keV. However, for {\it obs\#1}, where the MECS data are not available,
we used 0.5--5 keV and 5--25 keV for the LECS and HPGSPC, respectively. We
allowed for free normalization of the instruments in multi-instrument fits,
with respect to MECS in the HS (\S \ref{s:HS}) and the LECS in the SS (\S
\ref{s:SS}). For clarity of display, the spectra from multi-instrument fits
shown in Figures 3, 5 below were renormalized to the level of the MECS, and
LECS, respectively.

The quoted errors for the spectral parameters correspond to a 90\% confidence
level ($\Delta \chi^2 = 2.71$). Parameters in brackets in tables below were
fixed during fits. In modeling absorption and reflection, we assume the
abundances of Anders \& Ebihara (1982) and, following G99, the disk inclination
of $i=45\degr$.

\section{Results}
\label{results}

\subsection{Simple models to the LECS and MECS spectra}
\label{s:simple}

During the SS ({\it obs\#2}), the mean 2--10~keV flux level of the source was
$1.7 \times 10^{-8} \,$\ergcms, while during the HS ({\it obs\#3A}), the
corresponding flux was $7.9 \times 10^{-9} \,$\ergcms, i.e., lower by a factor
of 2. Our HS flux is higher than that typically measured in the HS by \asca,
$\sim 5\times 10^{-9} \,$\ergcms\ (E96), but it is similar, e.g., to the HS
value of
$7.5\times 10^{-9}$ erg cm$^{-2}$ s$^{-1}$ obtained for the observation 2 of
G97.

\begin{figure}
\epsscale{1.0}
\plotone{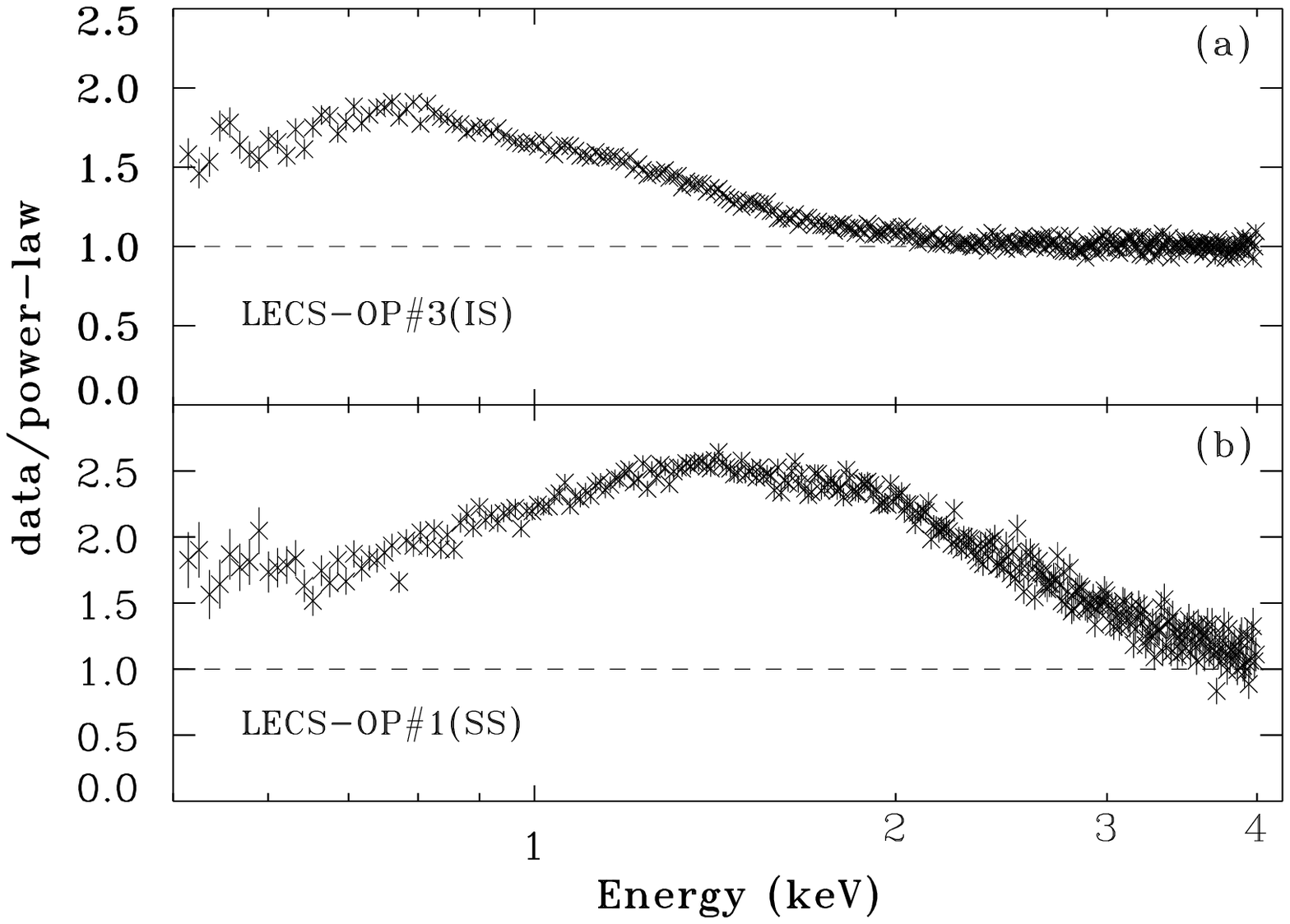}
\caption{Ratio of the LECS spectrum to the best-fitting power-law model
component (see Table 2) during the HS {\it (a)\/} and the SS {\it (b)}.
}
\label{f:comparison}
\end{figure}

We find that the energy range of each single NFI instrument is too narrow to
constrain physical models, in either of the observed states. A sum of a power
law (PL) and a blackbody (BB) photo-absorbed by a hydrogen column density,
$N_{\rm H}$, provides an acceptable fit (the reduced $\chi ^2 / \nu = 376/321$)
to the LECS (0.5--4 keV) continuum during the HS ({\it obs\#3A}). The same
model provides a very good fit ($\chi ^2 / \nu = 311/321$) to the corresponding
spectrum measured during the SS ({\it obs\#1}). In Table 2, we summarize the
results. As can be seen, we find a statistically significant higher value  of
the BB temperature in the SS, $kT_{\rm bb}\simeq 0.30$ keV, than $kT_{\rm
bb}\simeq 0.18$ keV found in the HS. Also, the PL slope is higher in the SS,
with the photon index of $\Gamma\simeq 2.8$, while $\Gamma\simeq 2.2$ in the
HS. Figure 2 compares the strength of the blackbody component in the HS and the
SS. The higher amplitude and temperature of the blackbody in the SS are
apparent. The hydrogen column density does not appear to significantly change
from {\it obs\#1} to {\it obs\#3A}, and thus also with the orbital phase
(see Table 1) of the binary period of Cyg X-1. This item will be discussed
later (\S 5.4).

\begin{deluxetable}{l c c}
\tabletypesize{\small}
\tablewidth{0pt}
\tablecaption{Parameters of the blackbody+power law model for the LECS 0.5--4
keV range}
\tablehead{
\colhead{Parameter} & \colhead{HS (obs.\ 3A)}  & \colhead{SS (obs.\ 1)} }
\startdata
$N_{\rm H}$ [$10^{21}$ cm$^{-2}$]   &  $5.4 \pm 0.3$   &   $5.9\pm 0.4$  \\
$\Gamma$        & $2.15  \pm 0.03$  & $2.80  \pm 0.15$  \\
$F_{\rm PL}(1\,{\rm keV})$ [cm$^{-2}$ s$^{-1}$] & $4.69\pm 0.04$  &
$17^{+17}_{-11}$ \\
$kT_{\rm bb}$ [keV]    & $0.184 \pm 0.006$ & $0.30 \pm 0.01$  \\
$L_{\rm bb}$ [$10^{37}$ erg s$^{-1}$] & $0.44  \pm 0.04$  & $2.3  \pm 0.4$ \\
$R_{\rm bb} $ [km]                      & $170   \pm 14$    & $147  \pm 15$ \\
$\chi ^2/\nu$              &      376/321      &       311/321  \\
\enddata
\end{deluxetable}

\begin{deluxetable}{l c c c c}
\tabletypesize{\small}
\tablewidth{0pt}
\tablecaption{Parameters of the blackbody+power-law+line model for the MECS
3--10 keV range}
\tablehead{
\multicolumn{1}{l}{Parameter} & \multicolumn{2}{c}{Gaussian line}
& \multicolumn{2}{c}{Disk line} \\
\colhead{}      & \colhead{HS (obs.\ 3A)} & \colhead{SS (obs.\ 2)} &
\colhead{HS (obs.\ 3A)} & \colhead{SS (obs.\ 2)}
}
\startdata
$N_{\rm H}$ [$10^{21}$ cm$^{-2}$]   &  [6.0]  &   [6.0] &  [6.0]  &   [6.0] \\
$\Gamma$        & $1.90 \pm 0.02$ & $2.78 \pm 0.03$ & $1.89 \pm 0.01$ & $2.70
\pm 0.03$ \\
$F_{\rm PL}(1\,{\rm keV})$ [cm$^{-2}$ s$^{-1}$] & $2.82 \pm 0.10$ & $10.5\pm
0.6$ & $2.75 \pm 0.08$ & $9.1\pm 0.6$ \\
$kT_{\rm bb}$ [keV]  & $0.276 \pm 0.1$ &
$0.36 \pm 0.01$ & $0.30 \pm 0.07$ & $0.37 \pm 0.01$ \\
$L_{\rm bb}$ [$10^{37}$ erg s$^{-1}$] & $0.4^{+3.6}_{-0.3}$ & $1.7 \pm
0.2$ & $0.3^{+1.2}_{-0.2}$ & $1.64 \pm 0.12$ \\
$E_{\rm Fe}$ [keV] &  $6.43 \pm 0.10$  & $6.54 \pm 0.04$ &  [6.4]  &  [6.4] \\
$\sigma_{\rm Fe}$ [keV] &  $0.68 \pm 0.16$  & $0.65 \pm 0.09$  & &      \\
$I_{\rm Fe}$ [$10^{-2}$ cm$^{-2}$ s$^{-1}$] &  $1.2 \pm 0.3$ & $1.65\pm 0.27$ &
$1.08
\pm 0.14$ & $1.55 \pm 0.15$ \\
EW [eV] & $146 \pm 40$  & $300 \pm 50$ &
$146 \pm 20$ & $300 \pm 30$  \\
$R_{\rm in}/R_{\rm g}$ & & &   $10^{+5}_{-4}$ &  $6^{+4}$  \\
$R_{\rm out}/R_{\rm g}$                     &     & & $100^{+90}_{-40}$ &
$100^{+50}_{-30}$     \\
$i$ [deg]      &    & & [45] & [45] \\
$\chi^2 /\nu$              & 175/143  & 160/143 & 169/143 & 199/143 \\
\enddata
\end{deluxetable}

A simple model (BB+PL) for the continuum plus a broad Gaussian line (with the
peak energy, width and photon flux of $E_{\rm Fe}$, $\sigma_{\rm Fe}$, and
$I_{\rm Fe}$, respectively) also provides a good fit ($\chi ^2 / \nu =
160/143$) to the 3--10 keV  spectrum measured by the MECS during the SS ({\it
obs\#2}).  We have also checked the effect of replacing the Gaussian line by a
disk line (DL) model of Fabian et al.\ (1989) with rest-frame energy $E_{\rm
Fe}$ and flux of $I_{\rm Fe}$, with the disk emitting the line from a radius,
$R_{\rm out}$, down to $R_{\rm in}\geq 6 R_{\rm g}$ (where $R_{\rm g}\equiv
GM/c^2$) with the surface emissivity dependence $\propto R^{-2}$. We have found
that the $\chi ^2 / \nu$ somewhat improves ($150/146$) for the inclination of
$i=60\degr$, but it gets significantly higher ($199/146$) for $i=45\degr$.
However, this worsening of the fit for $i=45\degr$ is no longer present in our
physical model (\S \ref{s:SS}). The best fit continuum parameters and the line
parameters for both models are reported in Table 3. Notice that the BB
temperature  estimated with MECS from {\it obs\#2} is slightly higher than that
obtained with LECS in the {\it obs\#1}. Given the MECS energy band, this higher
value is likely biased, and we consider the LECS result more reliable. The
values of $\Gamma$ estimated from {\it obs\#1\/} and {\it obs\#2\/} are instead
completely consistent within each other.

The BB+PL continuum model plus a Gaussian provides a marginally acceptable fit
also to the MECS 3--10 keV data in the HS ({\it obs\#3A}, see Table 3). The DL
profile provides a fit as good as the Gaussian. The best fit parameters of the
line are consistent with those obtained during the SS, except the line
equivalent width (EW), which is lower during the HS. The BB luminosity is
poorly constrained by the 3--10 keV data in the HS.

On the other hand, we find that simple  models fail to describe the broad band,
0.5--200 keV, spectra of {\it obs\#1} and {\it obs\#3B}. One can describe those
data in terms of Comptonization, blackbody emission, Compton reflection, and Fe
K$\alpha$ fluorescence, see \S \ref{s:HS} below.

\subsection{Physical model for the 0.5--200 keV spectrum of the HS}
\label{s:HS}

A power law, with Compton reflection and a Fe K line was found suitable to
describe the HS spectrum up to 20 keV (Palazzi et al.\ 1999). However, a simple
assumption of an exponential cutoff in the power law was insufficient to
describe the spectrum above 20 keV. Thus, we have investigated whether
replacing the e-folded power law by a spectrum from Comptonization can provide
a satisfactory description of the broad-band spectrum. We have found, however,
that none of an extensive array of models by either Poutanen \& Svensson (1996)
or Haardt (1993) for various geometries can, even approximately, describe the
actual data ({\it obs\#3B}).

On the other hand, we could obtain satisfactory fits constraining the energy
ranges either below or above $\sim 10$ keV. This provided us a motivation to
consider models with two Comptonizing regions, which are likely to provide a
better approximation to the actual X-ray source in Cyg X-1.  In support of this
scenario, G97 found that 2 thermal-Comptonization components were required to
account for the \ginga/OSSE hard-state data of Cyg X-1.

Thus, we have considered a model containing 2 thermal-Comptonization spectra,
corresponding to different temperatures $kT_1$ and $kT_2$, and the Thomson
optical depths, $\tau_1$ and $\tau_2$. For simplicity and to minimize the
number of free parameters, we have assumed that each of the Comptonization
sources can be represented by emission of a spherical plasma cloud with a
uniform distribution of seed blackbody photons (with $\tau$ measured along the
radius). We used the model {\tt compps} v3.4\footnote{{\tt compps} code is
available on the internet at ftp://ftp.astro.su.se/pub/juri/XSPEC/COMPPS}
(Poutanen \& Svensson 1996). As independent parameters, we choose $kT_1$,
$kT_2$ and the Compton parameters $y_1$ and $y_2$, where
\begin{equation}
y_i\equiv 4\tau_i {kT_i\over m_{\rm e}
c^2},
\end{equation}
and $i=1$, 2. This choice allows us to distinguish the spectral hardness of the
two components since a given value of $y$ corresponds to an approximately
constant value of $\Gamma$ (e.g., Ghisellini \& Haardt 1994; Poutanen 1998;
Beloborodov 1999b).

\begin{deluxetable}{l c}
\tabletypesize{\small}
\tablewidth{0pt}
\tablecaption{Parameters of the thermal-Compton model for the 0.5--200
keV HS spectrum (obs.\ 3B)}
\tablehead{\colhead{Parameter}      & \colhead{Value}  }
\startdata
$N_{\rm H}$ [$10^{21}$ cm$^{-2}$]     & [6.0]  \\
$kT_{\rm bb}$ [keV]         &  $0.16 \pm 0.02$  \\
$kT_1$ [keV]                      & $42 \pm 19$  \\
$y_1$                               & $0.15  \pm 0.01$    \\
$F_1 (1\,{\rm keV})$ [cm$^{-2}$ s$^{-1}$] & $1.7   \pm 0.8$    \\
$kT_2$ [keV]                   & $59 \pm 5$  \\
$y_2$                               & $0.89  \pm 0.01$    \\
$\Omega_2/2\pi$               & $0.25  \pm 0.04$   \\
$F_2 (1\,{\rm keV})$  [cm$^{-2}$ s$^{-1}$] & $1.55  \pm 0.09$    \\
$E_{\rm Fe}$ [keV]                            & $6.15  \pm 0.12$    \\
$\sigma_{\rm Fe}$ [keV]                      & $1.22  \pm 0.13$     \\
$I_{\rm Fe}$ [$10^{-2}$ cm$^{-2}$ s$^{-1}$] & $2.71 \pm 0.43$ \\
EW [eV]                                & $350 \pm 55$      \\
$\chi ^2 / \nu$                 & 377/361                   \\
\enddata
\end{deluxetable}

The present data are not able to constrain $N_{\rm H}$ with our model.
Therefore, it was fixed at the best-fit value obtained using the SS (see \S
\ref{s:SS}). Furthermore, the data do not allow us to distinguish between
rather similar components from Compton reflection of the emission of the two
plasma clouds. (For this process we use Green's functions of Magdziarz \&
Zdziarski 1995.) Therefore, we have exploited here a correlation between the
relative strength of Compton reflection, $\Omega/2\pi$, and $\Gamma$ found in
Cyg X-1 by Gilfanov, Churazov \& Revnivtsev  (1999, hereafter GCR99). This
correlation is also present in other black-hole binaries as well as in
Seyfert-1 galaxies (Zdziarski, Lubi\'nski \& Smith 1999). According to it, the
softer the spectrum, the larger the value of $\Omega/2\pi$. To take into
account this effect, we have assumed that $\Omega/2\pi$ of the softer component
is twice that of the harder component. The seed photons for Comptonization are
assumed to have a blackbody distribution with a temperature, $kT_{\rm bb}$.

This  model does indeed provide a good description of the data, see Table 4.
From equation (1), the optical depths corresponding to the fitted values of $y$
of the soft and hard Comptonization components are $\tau_1 = 0.5 \pm 0.2$ and
$\tau_2 = 1.9 \pm 0.2 $, respectively. No additional (i.e., not passing through
the hot plasma) blackbody component is required by the fit, but its presence
cannot be ruled out. We also find that replacing the blackbody distribution of
the seed photons by that of a disk blackbody yields a fit worse by
$\Delta\chi^2 =+6.3$.

Figure 3 compares the resulting model spectrum with the data. We see that the
hard component dominates at energies $\ga 1.5$ keV, with the soft one providing
a soft excess, dominant at lower energies. The temperature of the soft
component is thus weakly constrained due to its small contribution to the
spectrum around the high-energy cutoff. Compton reflection is dominated by that
of the harder component, with $\Omega_2/2\pi =0.25 \pm 0.04$.

In addition to the continuum, a strong and broad line is also present in the
spectrum. The residuals of the data to the continuum model are shown in Figure
4. The line and edge region are much better fitted by static reflection and a
Gaussian ($\chi ^2 /\nu = 377/361$, see Table 4) than by the line and
reflection coming from a disk. In the latter case, best fits have been obtained
for the radial dependence of the line and reflection being the same as that of
the disk emission itself. Still, the obtained fits are much worse than that of
Table 4, $\chi ^2 / \nu = 469/362$ and 451/362 for $i=45\degr$ and $60\degr$,
respectively. At $i=60\degr$, EW $=190$ eV and $\Omega_2/2\pi=0.42$.

\begin{figure}
\psfig{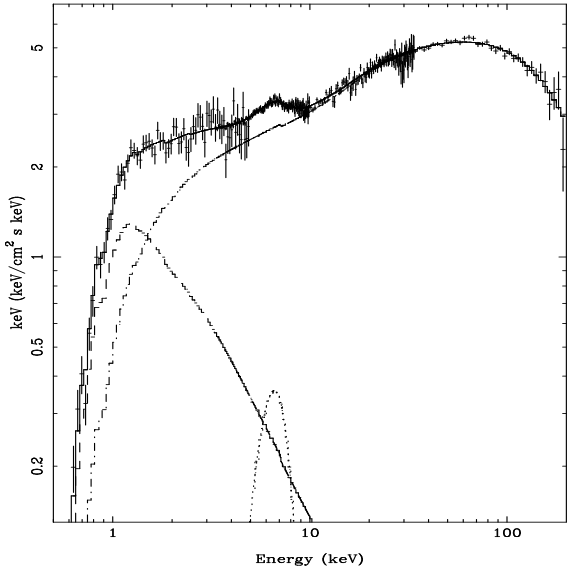}
\caption{The broad-band $EF_E$ spectrum during the HS (crosses). The histograms
show the model components due to 2 thermal-Compton plasmas with Compton
reflection and a broad Gaussian (see \S \ref{s:HS} and Table 4).
}
\label{f:HS}
\end{figure}

\begin{figure}
\psfig{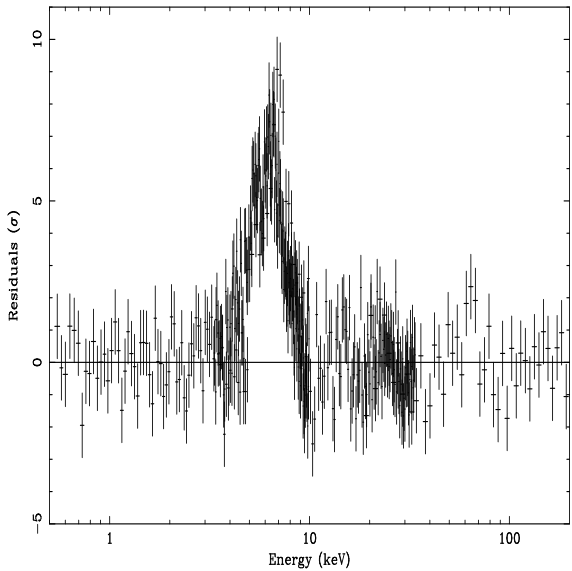}
\caption{The ratio of the HS data to the best-fit continuum model (Fig.\ 3 and
Table 4) excluding  the Fe K$\alpha$ line, which large width is apparent.
}
\label{f:line}
\end{figure}

The total unabsorbed 0.5--200 keV flux and the bolometric flux derived from our
model are $4.2\times 10^{-8}$ erg cm$^{- 2}$ s$^{-1}$ and $5.1\times 10^{-8}$
erg cm$^{- 2}$ s$^{-1}$ (corresponding to the luminosities of $\sim 2.0\times
10^{37}$ and $2.4\times 10^{37}$ erg s$^{-1}$ at $d=2$ kpc and assuming
isotropy), respectively. The bolometric fluxes in the hard and soft components
are $3.5\times 10^{- 8}$ erg cm$^{-2}$ s$^{-1}$ and $1.6\times 10^{- 8}$ erg
cm$^{- 2}$ s$^{-1}$, respectively.

\subsection{Physical model for the 0.5--200 keV spectrum in the SS}
\label{s:SS}

The above thermal-Comptonization model does not fit the SS spectrum. The
reasons are the same as those given by G99 for the SS measured by \asca/\xte\/
1996 May 30 (who also took into account the OSSE data of 1996 June 14--25);
namely, the presence of an extended power-law like tail at high energies, and
of a soft excess. G99 have fitted their data by a non-thermal Comptonization
model developed from a code of Coppi (1992), see also Poutanen \& Coppi (1998).

The model is described in G99, where we refer the reader for details. Its main
ingredient is a hot plasma cloud with continuous acceleration of electrons at a
rate $\propto \gamma^{-\Gamma_{\rm inj}}$. The high-energy electrons lose
energy due to Compton, Coulomb and bremsstrahlung processes, and thus establish
a steady-state distribution. At high energies, the distribution is non-thermal
(power-law like), but at low energies a Maxwellian distribution is established.
The temperature of the Maxwellian, $kT_{\rm e}$, is determined by balance
between Compton gains and losses, Coulomb heating by high-energy electrons,
bremsstrahlung losses, and a direct heating (e.g., Coulomb heating by energetic
ions). The total number of Maxwellian electrons (not including e$^\pm$ pairs,
the production of which is also taken into account) is determined by the
corresponding Thomson optical depth, $\tau_{\rm i}$ (the second free
parameter). The cloud is irradiated by blackbody photons emitted by an
accretion disk. These photons serve as seed for Compton scattering by both
thermal and non-thermal electrons.

The system is characterized by powers, $L_i$, supplied to its different
components. We express each of them dimensionlesslly as a compactness, $\ell_i
\equiv L_i \sigma_{\rm T}/({\cal R} m_{\rm e} c^3 $), ${\cal R}$ is the
characteristic dimension of the plasma and $\sigma_{\rm T}$ is the Thomson
cross section. $\ell_{\rm s}$, $\ell_{\rm th}$, $\ell_{\rm nth}$, and
$\ell_{\rm h}= \ell_{\rm th}+\ell_{\rm nth}$ correspond to the power in soft
disk photons irradiating the plasma, in direct electron heating, in electron
acceleration, and total power supplied to electrons in the plasma,
respectively. As found by G99, the fits allow a range of $\ell_{\rm s}$, in
which the dominant energy loss process of energetic electrons is Compton, and
e$^\pm$ pair production does not yield a visible annihilation feature (not seen
in the OSSE data). Following G99, we fix here $\ell_{\rm s}=10$.

The disk spectrum incident on the plasma is modeled as coming from a
pseudo-Newtonian accretion disk extending from infinity down to the minimum
stable orbit, $R_{\rm in}= 6R_{\rm g}$, the spectral shape of which can be
characterized by the maximum color temperature of the disk, $kT_{\rm max}$. The
spectrum from both reflection and the Fe K$\alpha$ fluorescence is calculated
taking into account relativistic smearing with $R_{\rm out}=10^3 R_{\rm g}$,
$R_{\rm in}$ as above, and the emissivity dependence $\propto R^{-2}$ (see \S
\ref{s:simple}). The reflected surface is allowed to be ionized, with the
degree of ionization characterized by the ionization parameter, $\xi\equiv 4
\pi F_{\rm ion}/n$ (where $F_{\rm ion}$ is the ionizing flux and $n$ is the
reflector density).

The fit does not change with the disk inclination ($\chi^2 /\nu =167.9/184$,
167.1/184 for $i=45\degr$, $60\degr$, respectively). The best-fit parameters
for $i=45\degr$ are given in Table 5, which, for comparison, also shows the
results of G99. Figure 5 shows the observed broad band spectrum as well as the
model and its components. Note that we find a lower disk ionization than that
in the observation of G99.

\begin{figure}
\psfig{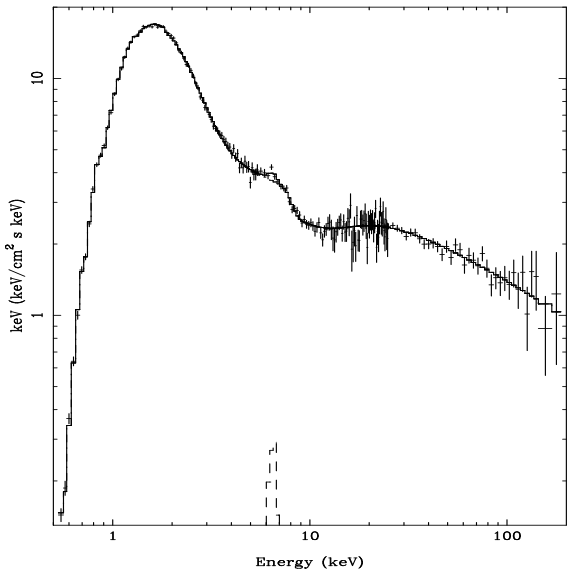}
\caption{The broad-band $EF_E$ spectrum during the SS (crosses). The histograms
show the continuum model with the hybrid plasma Compton upscattering blackbody
disk photons, and a broad disk line.}
\label{f:SS}
\end{figure}

\begin{deluxetable}{l c c}
\tabletypesize{\small}
\tablewidth{0pt}
\tablecaption{Parameters of the hybrid model for the 0.5--200 keV SS spectrum
(obs.\ 1)}
\tablehead{
\colhead{Parameter} &
\colhead{\sax} &
\colhead{\asca/\xte}\\
\colhead{ } &
\colhead{(June 22)} &
\colhead{(May 30)}
}
\startdata
$N_{\rm H}$ [$10^{21}$ cm$^{-2}$] & $6.0   \pm 0.1$  &  $5.2^{+0.1}_{-0.2}$ \\
$kT_{\rm max}$ [keV]   & $0.368 \pm 0.006$       & $0.36^{+0.01}_{-0.01}$ \\
$\ell_{\rm h}/\ell_{\rm s}$ & $0.168^{+0.015}_{-0.020}$  &
$0.35^{+0.01}_{-0.02}$
\\
$\ell_{\rm nth}/\ell_{\rm h}$    & $0.6\pm 0.20$ & $0.77^{+0.05}_{-0.04}$
\\
$\tau_{\rm i}$  & $0.098^{+0.016}_{-0.012} $  & $0.25^{+0.05}_{-0.04}$  \\
$\Gamma_{\rm inj}$   & $2.4  \pm 0.2$   & $2.54^{+0.11}_{-0.07}$  \\
$\Omega/2\pi$   & $1.5^{+1.4}_{-0.6}$   & $0.63^{+0.13}_{-0.11}$   \\
$\xi$ [erg cm s$^{-1}$] & $65^{+33}_{-23}$ & 350$^{+250}_{-120}$   \\
$E_{\rm Fe}$ [keV]   & [6.4]     & $6.37^{+0.14}_{-0.14}$   \\
EW [eV]  & $80 \pm 60$             & $120^{+50}_{-30}$ \\
$\chi^2 / \nu$             &  167.9/184               & 618/574 \\
\enddata
\end{deluxetable}

The  0.5--200 keV and bolometric unabsorbed fluxes (normalized to the LECS) are 
$1.0\times 10^{-7} \,$ erg cm$^{-2}$ s$^{-1}$ and $1.5\times 10^{-7} \,$ erg
cm$^{-2}$ s$^{-1}$ ($L\simeq 7.1 \times 10^{37}$ erg s$^{-1}$ assuming
isotropy), respectively. They are higher by a factor of $\sim 3$ than those in
the HS. The main change in the flux is in the 0.5--2 keV range, where the
blackbody photon flux is more prominent; $1.3\times 10^{-8} \,$\ergcms\ in the
HS versus $7.7 \times 10^{-8} \,$\ergcms\ in the SS.

So far, we have modeled the Fe K$\alpha$ line as coming from a disk extending
down to $R_{\rm in}= 6R_{\rm g}$. When we relax this assumption, we obtain
$R_{\rm in}=6^{+5}_{-0} R_{\rm g}$. Alternatively, the line can be described as
a Gaussian, with $\chi^2 /\nu=165/183$, very similar to the disk-line case.
Then, $E_{\rm Fe}= 6.2\pm 0.3$~keV and $\sigma_{\rm Fe} =0.4\pm 0.4$ keV.

We notice that the (single-component) hybrid model is not suitable to describe
the 0.5--200~keV HS spectrum. Even if we restrict the energy band to
4--200~keV, which is dominated by the hard Comptonization component in our
spectrum, we find a rather unsatisfactory fit ($\chi^2 /\nu = 374/281$),
corresponding to a probability of only $1.6 \times 10^{-4}$ that this model is
correct. This shows that the spectrum observed by us (belonging to the HS, see
\S 1) is different from the spectra observed by \xte\/ during the spectral
transition on 1996 May 22--23, which were successfully fitted by G99 with the
hybrid model in the same energy range.

\section{Discussion}
\label{discussion}

\subsection{Accretion flow in the HS}
\label{accretion:HS}

One of our main new results is the interpretation of the soft excess in the HS
state as being due to thermal Comptonization by a hot plasma with a low value
of the Compton parameter ($y$), whereas the main, hard X-ray, continuum is due
to thermal Comptonization by a plasma with a higher $y$. Our model with two
plasma components represents, most likely, an approximation of a spatial
distribution of the temperature and optical depth with radius.

This result appears to apply to the HS in general, with a soft excess at $\la
3$ keV being commonly found (E96, G97). In fact, a strong evidence for a radial
stratification of the X-ray source in Cyg X-1 in the HS is provided by results
of Fourier-frequency resolved spectroscopy (Revnivtsev, Gilfanov \& Churazov
1999; GCR99). These authors show that the total spectrum of Cyg X-1 contains
components with different values of $\Gamma$ corresponding to different
frequency ranges of temporal variability.

Figure 3 shows that the hard component dominates at energies $\ga 1.5$ keV and
the bolometric emission (\S \ref{s:HS}). This is, in fact, in agreement with
the result of GCR99, who found that the time-averaged spectrum in the HS of Cyg
X-1 is similar to the hardest of the frequency-resolved spectra (see their
fig.\ 8). This implies that the average spectrum is dominated by a hard
component constant in time, similar to the situation in our model.

The temperature of the main, hard component, $kT\simeq 60$ keV, is less than
that measured by G97 ($kT\sim 100$ keV). On the other hand, our value of $kT$
is almost the same as that found in the HS of another persistent black-hole
binary, GX 339--4 (Zdziarski et al.\ 1998). The temperature of our soft
component is similar, but not well constrained due to its small contribution at
highest energies (see Fig.\ 3). The plasma emitting the soft component appears
to be farther away from the black hole according to the decrease of the typical
frequency of the intensity variations with $\Gamma$ in the spectroscopy of
Revnivtsev et al.\ (1999). This is also supported by our result that the
Thomson optical depth of the softer component is much less than that of the
hard one, in agreement with accretion disk models, which generic feature is an
increase of $\tau$ inward. We note that our model of the soft excess is
different from that in which the soft emission is attributed to Comptonization
in a warm (with $kT\sim 1$ keV) surface layer above the accretion disk (e.g.,
Czerny \& Elvis 1987; Zhang et al.\ 2000).

We note that our result that the line in the HS is broad agrees with results of
Done \& \.Zycki (1999) for the HS, who found broad lines in their reanalysis of
the \asca\/ and {\it EXOSAT\/} data, even though from previous studies of
those data (Done et al.\ 1992; E96) the lines were found to be narrow. Also,
broad lines were reported in the \xte\/ data (GCR99).

An important unresolved issue for our model is the strength of the Fe K$\alpha$
line as compared to the strength of reflection (dominated by the hard
component). Namely, as seen in Table 4, the relatively low reflection strength
of $\Omega/2\pi\sim 0.25$ cannot explain the strong Fe K$\alpha$ line with EW
of $\sim 350$ eV (see, e.g., \.Zycki \& Czerny 1994). We feel that this
disagreement is due to our two-zone approximation for the incident continuum
being insufficient to describe the actual distribution of plasma parameters.
The strong dependence on the assumed continuum is shown by comparison with the
results assuming relativistic smearing by a disk, in which case EW $\simeq 190$
eV and $\Omega/2\pi\simeq 0.42$ (\S \ref{s:HS}) as well as with Table 3, in
which we found EW $\simeq 150$ eV assuming a power-law continuum for the MECS
data only. Furthermore, our finding that a Gaussian line fits the data better
than the disk line (\S \ref{s:HS}) probably indicates a distribution of the
cold medium different from a disk, e.g., blobs embedded in a hot medium
(Poutanen 1998; Zdziarski et al.\ 1998; B\"ottcher \& Liang 1999) or an
outflow. 

In principle, it is possible that our obtained low value of the reflection
strength is due to heating and photoionization of a surface layer of the
accretion disk (e.g., Nayakshin 2000; Nayakshin, Kazanas \& Kallman 2000).
However, we find no signature of ionization in the data (the best-fit $\xi=0$).
Furthermore, the strength of reflection is clearly higher in the SS, contrary
to an expectation of that model that a much higher soft X-ray ionizing flux
would reduce reflection even further.

We also note that the value of $\Omega$ is less than that expected
on the basis of the $\Omega(\Gamma)$ correlation of GCR99. This appears to be,
at least partly, due to using different incident continuum models by us and by
GCR99, who  used spectra below 20 keV and the {\tt pexrav} model of {\sc
xspec} for their fitting. In fact, our spectral data fitted below 20 keV with
the same model (Palazzi et al.\ 1999) are in agreement with the
correlation of GCR99.

\subsection{Accretion flow in the SS}
\label{accretion:SS}

Our results (\S \ref{s:SS}) strongly confirm the applicability of the hybrid
plasma model to the SS, as found by G99. Those authors present an exhaustive
discussion of physical consequences of this model, which we thus do not repeat
here. Most of the model parameters obtained by us and by G99 for their
\asca/\xte\/ observation are rather similar, as shown in Table 5. However, when
$R_{\rm in}$ is allowed to be free, G99 find its value as $17^{+26}_{-8} R_{\rm
g}$, whereas we obtain $6^{+5}_{-0} R_{\rm g}$. This appears to indicate that
the inner radius of the disk moved down to the minimum stable orbit between
1996 May 30 and June 22. This is also supported by the larger solid angle of
the reflector obtained for the observation on June 22, $\Omega/2\pi= 1.5\pm
0.6$ as compared to $0.63^{+0.13}_{-0.11}$ on May 30. Indeed, the temporal and
spectral properties of Cyg X-1 on May 30 indicate the object was still
undergoing a transition between the HS and SS (Cui et al.\ 1997; Pottschmidt et
al.\ 2000; Wen et al.\ 2000).

We note that the EW of the Fe K$\alpha$ line obtained by us, $80\pm 60$ eV, is
marginally too low for the strength of reflection with $\Omega/2\pi= 1.5\pm
0.6$ (\.Zycki \& Czerny 1994). Given that  our line profile relies on the data
from HPGSPC (as MECS were not usable), we have also checked the EW of the line
in {\it obs\#2}, when MECS were switched on. With the same continuum model we
found a similar value of the EW of $90\pm 40$ eV. Still, the values of $\Omega$
and EW are consistent with each other within the 90\% uncertainty ranges.

Other physical processes proposed to dominate in the SS are thermal
Comptonization (e.g., Poutanen, Krolik \& Ryde 1997; Cui et al.\ 1998; Esin et
al.\ 1998) and bulk-motion Comptonization (e.g., Laurent \& Titarchuk 1999).
Here, we have not been able to fit the \sax\/ SS data with thermal
Comptonization, confirming previous ruling out of this model (G99; Zdziarski
2000).

The bulk-motion Comptonization model is ruled out on the basis of the lack of a
sharp high-energy cutoff at $\sim 100$--200 keV (predicted by that model,
Laurent \& Titarchuk 1999) and the very large accretion rate required to
account for the relatively hard power-law tail in the SS spectrum of Cyg X-1
(see Zdziarski 2000 for detailed discussion). We also note that McConnell et
al.\ (2000) find a strong 1--10 MeV flux from Cyg X-1 in the SS observed by the
\gro/COMPTEL. That flux is completely inconsistent with the predictions of the
bulk-Compton model. On the other hand, our best-fit hybrid model for the SS
predicts no high-energy cutoff up to 10 MeV, and we find the predictions of
that model for the 1--10 MeV range to be in agreement with the SS measurements
of McConnell et al.\ (2000).

\subsection{The nature of state transitions}
\label{s:state}

In \S\S \ref{s:HS}--\ref{s:SS}, we have calculated that the bolometric fluxes
(extrapolated from the 0.5--200 keV range using best-fitting models) in the HS
and SS were $\sim 5\times 10^{-8}$ erg cm$^{-2}$ s$^{-1}$ and $1.5\times
10^{-7}$~erg cm$^{-2}$ s$^{-1}$, respectively. On the other hand, we have found
that the bolometric flux during 4 observations of Cyg X-1 in 1991 in the HS by
\ginga\/ and OSSE (G97) was within the range of $\sim (3$--$6)\times 10^{-8}
\,$\ergcms. The bolometric flux for the SS \asca/\xte\/ data of G99 was
$1.3\times 10^{-7} \,$\ergcms\ (normalized to the PCA). Thus, the bolometric
flux measured by us in the 2 states appear typical. Our value of the bolometric
flux in the SS is slightly higher than that found for the 1996 May 30
observation by G99, which may be caused by Cyg X-1 not yet having fully reached
its SS during that observation.

Thus, we find that the increase of the bolometric flux in the SS with respect
to that in the HS is by a factor of $\sim 3$. On the other hand, Zhang et al.\
(1997a) claimed the 1.3--200 keV flux remaining almost unchanged (within $\sim
15\%$) during transitions from the HS to SS, as well as constrained the change
of the bolometric flux to be $\la 1.5$. The disagreement between those results
and ours probably stems from the uncertain relative calibration of the \xte/ASM
and \gro/BATSE instruments used by Zhang et al.\ (1997a), and the fact the bulk
of the flux in the HS and SS falls in the energy range of BATSE and ASM,
respectively. On the other hand, we present here the flux measurements in the 2
states with the same instruments and with a broad-band energy coverage.

If we assume isotropy, the (bolometric) luminosities in the HS and SS measured
by us are $L =0.016 (d/{\rm 2\, kpc})^2 (M/10 M_\odot) L_{\rm E}$ and $L=0.048
(d/{\rm 2\, kpc})^2 (M/10 M_\odot) L_{\rm E}$, respectively. Here, $L_{\rm E}$
$\equiv 4 \pi \mu_{\rm e} G M c m_{\rm p}/\sigma_{\rm T}$ is the Eddington
luminosity, $\sigma_{\rm T}$ is the Thomson cross section, $\mu_{\rm e} =
2/(1 + X)$ is the mean electron molecular weight and $X \approx 0.7$ is the
hydrogen mass fraction. For $X=0.7$, $L_{\rm E}\simeq 1.48\times 10^{38}
(M/M_\odot)$ erg s$^{-1}$.

However, the SS-to-HS luminosity ratio can be different than the flux ratio
(G99, Beloborodov 1999a). One reason for that is that while the HS emission,
coming mostly from an optically-thin plasma, may be isotropic, the SS emission
is predominantly due to the anisotropic blackbody radiation. If $i<60\degr$
(which is probably the case in Cyg X-1), the observed disk flux will be higher
than that coming from an isotropic source with the same $L$. For the
observations of G99 and $i=35\degr$, the reduction of $L$ due to this effect is
by about 3/4, which would correspondingly reduce the luminosity ratio between
the SS and the HS. On the other hand, the hot plasma emission in the HS can be
beamed towards the observer (Beloborodov 1999a), and thus anisotropic as well,
in which case the luminosity ratio can be higher than the observed flux ratio.

The low relative reflection obtained during the HS (Table 4) would be in favor
of a large inner radius of the accretion disk during this state. A large
$R_{\rm in}$ is also in agreement with the BB flux and temperature lower than
those in SS, but it is in conflict with the width, peak energy and EW ($\sim
1.2$~keV, $\sim 6.2$~keV and 350~eV, respectively) of the Fe K$\alpha$ line
observed during HS.  On the other hand, the accretion flow in the SS is clearly
dominated by an optically-thick accretion disk extending down to the minimum
stable orbit (\S \ref{s:SS}).

Any theory of the state transition in Cyg X-1 has to take into account the
above results. One possible cause of the transition is a change of $\dot M$. We
note here that our above results on the SS-to-HS luminosity ratio being
significantly larger than the previously-found value of $\la 1.5$ (and likely
as large as 3) make models based on the change of $\dot M$ much more plausible,
as requiring now much less fine-tuning.

The most detailed model postulating a change of $\dot M$ is probably that of
Esin et al.\ (1998), in which the HS is very close to the maximum possible
$\dot M$ of an optically-thin advective flow (Narayan \& Yi 1995; Abramowicz et
al.\ 1995), and a further increase leads to a switch to accretion dominated by
an optically-thick accretion (Shakura \& Sunyaev 1973). We note that the model
in its original version postulated that an optically-thick flow in the HS is
truncated at $\ga 10^3 R_{\rm g}$. This  appears to be in conflict with the Fe
K line properties discussed above and the relativistic smearing also observed
in the HS state (Done \& \.Zycki 1999). Also, another important modification
required in that model is that the corona above the disk in the SS has to be
non-thermal.

A related model attributes the state transitions to evaporation of the inner
disk, which physical process occurs at a low $\dot M$ (Meyer, Liu \&
Meyer-Hofmeister 2000; R\'o\.za\'nska \& Czerny 2000). Thus, the innermost flow
in the HS consists of a hot plasma only. Then, a cold accretion disk with and
without a hot corona exists at intermediate and large radii, respectively. The
Fe K$\alpha$ line is then formed at the intermediate radii. In the SS, the cold
accretion disk extends down to the minimum stable orbit.

Beloborodov (1999a) has proposed that the source in the HS does not form
an advective hot disk but, rather, a mildly relativistic outflow above the
surface of an optically-thick disk. The HS-to-SS transition would then involve
the corona becoming static as well as non-thermal. The data studied here appear
not to allow distinguishing between these possibilities.

Finally, Zhang, Cui \& Chen (1997b) have proposed that the Cyg X-1 state
transition is due to reversal of the direction of disk rotation from retrograde
(HS) to prograde with the black hole spinning with an angular momentum of $0.75
GM/c$. Disk reversals are possible from accretion fluctuations in stellar wind
\cite{Shapiro76}. Even if the transition to the HS was not terminated during
our observations, some of the consequences discussed by Zhang, Cui \& Chen
(1997b) could be in agreement with our results and with the observed variation
of the total X-ray luminosity of Cyg X-1 during transition from SS to HS.
However, this model has several difficulties. One is that the state transition
in Cyg X- 1, accreting from a focused wind, should be caused by something
different than the state transition in low-mass X-ray binaries, accreting from
a Roche-lobe overflow. Then, it is unclear why the focused-wind accretion
should be preferentially retrograde (with Cyg X-1 spending $\sim 90\%$ of time
in the HS). We also note that the relationship between emission of the hot
plasma and of the cold disk in the 2 states remains unexplained in this model.

\subsection{Hydrogen column density}
\label{NH}

The hydrogen column density does not appear to significantly change with the
source spectral state in our observations. Using the BB+PL model for the LECS
data alone (Table 2) in the HS and SS, we find $N_{\rm H} = (5.4\pm 0.3)$ and
$(5.9\pm 0.4)\times 10^{21}$ cm$^{-2}$, respectively. The latter value fully
agrees with that obtained using the hybrid Compton model for the SS broad-band
spectrum (\S \ref{s:SS}), $(6.0\pm 0.1) \times 10^{21}$ cm$^{-2}$.

We can compare those results with those based on the reddening of the system,
which has been measured as $E(B-V)=1.12\pm 0.05$ (Bregman et al.\ 1973) and
$0.95\pm 0.07$ (Wu et al.\ 1982). The 2 values were obtained by averaging the
reddening over regions centered on Cyg X-1 with the sizes of $30'$ and 
$7\degr$, respectively. No variation of the reddening with the orbital phase of
the binary system was found (Wu et al.\ 1982). From the most extended all-sky
study of the distribution of neutral H based on high-resolution {\it IUE\/}
observations of Ly$\alpha$ absorption towards 554 OB stars it has been found
that their $N_{\rm H}$ is well correlated with the column density of dust,
measured by $E(B-V)$, with $\langle N_{\rm H}/E(B-V) \rangle = 4.93 \times
10^{21}\, {\rm cm^{-2}\, mag^{-1}}$ (Diplas \& Savage 1994). This implies
$N_{\rm H}= (5.5\pm 0.2)\times 10^{21}\, {\rm cm^{-2}}$ and $(4.7\pm 0.3)\times
10^{21}\, {\rm cm^{-2}}$ for the estimates of the extinction of Bregman et al.\
(1973) and Wu et al.\ (1982), respectively.

Our values of $N_{\rm H}$, unlike the estimates based on the reddening, are
also sensitive to the effect of absorption by circumstellar matter within the
system, in which accretion proceeds via a focused wind from the companion. The
wind is partially ionized and its observed column depends on the orbital phase,
which leads to a modulation of the X-ray flux with the phase, discovered in the
ASM data by Wen et al.\ (1999). Thus, we could have expected a significant
difference between $N_{\rm H}$ in our HS observations (phase 0.51--0.55) and
the SS one (phase 0.93--0.94). However, Wen et al.\ (1999) have found that the
modulation becomes much weaker in the SS, which is, most likely, explained by
the stronger ionization of the wind by the stronger soft X-ray flux in that
state. This can explain our value of $N_{\rm H}$ at the phase $\sim 0$ in the
SS being compatible with the value at the phase $\sim 0.5$ in the HS.
Furthermore, Ba{\l}uci\'nska-Church et al.\ (2000) have attributed the
modulation to the effects of blobs of neutral matter causing the X-ray dips
observed preferentially close to the phase 0. In our data, we have not seen any
X-ray dips, which is consistent with the near uniformity of our measured values
of $N_{\rm H}$.

\acknowledgements

AAZ has been supported in part by KBN grants 2P03C00511p0(1,4) and 2P03D00614
and a grant from the Foundation for Polish Science. We thank M. Gierli\'nski
for his assistance with the {\sc xspec} software package, M. Guainazzi for his
support with the LECS data analysis, J. Poutanen for comments on this paper,
and G. Matt and S. N. Zhang for useful discussions. The \sax\/ program is
supported by the Italian Space Agency (ASI).

 \end{document}